# The application of Nano-silica gel in sealing well micro-annuli and cement channeling

Olatunji Olayiwola[1], Vu Nguyen[1], Randy Andres[1], Ning Liu[1].

[1] University of Louisiana at Lafayette, USA

**Abstract**

The possibility for hydrocarbon fluids to migrate through debonded micro-annuli wells is a major concern in the petroleum industry. With effective permeability of 0.1–1.0 mD, the existence of channels in a cement annulus with apertures of 10–300 µm constitutes a major threat. Squeeze cement is typically difficult to repair channels-leakage with small apertures; hence, a low-viscosity sealer that can be inserted into these channels while producing a long-term resilient seal is sought. A novel application using nano-silica sealants could be the key to seal these channels. In the construction and sealing of hydrocarbon wells, cementing is a critical phase. Cement is prone to cracking during the life cycle of a well because of the changes in downhole conditions. The usage of micro-sized cross-linked nano-silica gel as a sealant material to minimize damaged cement sheaths is investigated in this study. Fluid leakage through channels in the cement was investigated using an experimental system. With a diameter of 0.05 inches, the impact of the cement channel size was explored. The sealing efficiency increased from 86 percent to 95 percent when the nano-silica concentration of the sealing gel increased from 13 percent to 25 percent. This demonstrates that the concentration of nano-silica in the sealing gel affects the gel's ability to seal against fluid flow. This research proposes a new way for improving cement zonal isolation and thereby lowering the impact of cement failure in the oil and gas industry.

**Introduction**

Cementing engineering is one of the most important issues in the drilling operations, which operates in tough and challenging settings with temperature and pressure restrictions (Wojtanowicz et al., 2016). When a well fails, various factors come into play, including leakage, cementing, and so on. In this scenario, a variety of solutions can successfully improve industrial processes, with a focus on cement characteristics, crack sealing, shielding permeability reduction, gas migration management, and so on. During hydrocarbon production, external influences such as temperature and pressure cause the well cement to fracture, allowing fluid to leak at the wellbore contact (Dusseault et al, 2014). As a result, the leak has a substantial influence on the ecology, putting oil workers' and other aquatic wildlife's safety in jeopardy (Davies et al, 2014).

Cement is prone to cracking during the life of the well because of changes in downhole environment. During drilling or production operations, the cement integrity must never be jeopardized. If the cement is not properly finished and abandoned, leaks can occur at any time during the well's life (Watson and Bachu, 2009). Fluids (water or hydrocarbons) can move through channels within the cement or between the cement and its environment. These channels form when the wellbore integrity is compromised, allowing formation fluids to travel between formations and/or from the formation to the surface. Personnel and the environment are put at risk when water, gases, and hydrocarbon substances leak through cement routes (Davies et al., 2014).

Mechanical failures as a result of pressure and temperature cycles, chemical degradation owing to corrosive formation fluids (Zhang and Bachu, 2010), or inappropriate slurry design can all provide leakage paths in the cement annulus. The use of low-density slurries, which may allow formation fluids to create channels within the cement sheath, high-fluid-loss slurries, which may affect the mechanical properties of the set cement, and rigid cement, which can get fracked as a result of casing expansions and contractions, are all examples of poor slurry design. Furthermore, failures might arise as a result of incorrect mud clearance. The mud on the formation's and casing's surface can hinder the cement from adhering with its surroundings, resulting in micro-annuli. These and other defects can occur, allowing formation fluids to pass through the cement barrier.

Cement squeeze remedial operations are commonly used to fix cement leaks. However, due to low cement injectivity, pressure restriction, pin-hole leakage, micro-channels and fractures inside the leaked formation, and micro-cracks within primary cement, cement squeeze cannot be employed successfully to reduce and prevent leakage in some instances (Jones et al. 2014). Although micro fine cement is designed to promote penetration, the underlying problem of cement and steel casing adhesion versus cohesive bond failure in the downhole environment is not addressed by most, if not all, current repair materials. Adhesion failure is characterized by brittleness and unpredictability. A typical adhesion failure is cement-steel failure.

However, delivering the gel fluid to the fracture zone before it gels is a major challenge in remedial operations. The rheology of the gel controls the occurrence. (Zareie, C., 2019). To address this problem, the gel mixture's composition is designed to have a long gelation time and low viscosity, allowing it to be pumped and crosslinked during oil well workovers (Zhang, X. et. al, 2015). Because of their low thermal conductivity, low refractive index, high optical transparency, and large surface area, Nano-silica particles have gotten a lot of attention in well remedial operations. Cross-linking of polymers contributes to the formation of the silica network. The gel state, a colloidal intermediate phase between the solution and the solid state, has a three-dimensional structure and has promising applications in a variety of fields.

**State of the Art**

Cement squeeze operations are used to repair wells with poor primary cementing jobs or suspected leaks, in which additional cement is injected through perforations produced in the casing near the suspected source of leakage to ensure appropriate zonal isolation. Squeeze cementing is a corrective procedure that includes applying differential pressure across the cement slurry in order to dehydrate the cement (Goodwin, 1984). In theory, the slurry is supposed to reach and fill the problem area, immobilizing the area until compressive strength can be developed.

However, because to a miscalculation of the leakage problem, cement slurry is frequently inadequately positioned or planned. Even if an annular gap and/or cracks with apertures on the order of 0.01–0.3 microns are present, effective permeability in the region of 0.1–1 mD can be significantly increased (Jung et al., 2014). Fractures or leakage paths with small apertures, in particular, are difficult to repair using ordinary Portland cement because the cement slurry is screened out of the dispersion fluid and cannot penetrate the fracture. As a result, squeeze cementing is frequently unsuccessful, resulting in rig time waste and excessive expenses.

The particle size is the key factor limiting the sealing effectiveness of traditional Portland cement. Class H oil well cement comprises large particles in the range of 100-150 micrometres, making it difficult to penetrate small channels, micro-annuli, or thin mud channels, and often resulting in unsatisfactory results. When the slurry is squeezed to penetrate fractures thinner than 400 micrometres, bridging and cement dehydration will occur. Ultra-fine cement technologies have been developed to reduce the particle size of cement slurries, such as Halliburton's Micro Matrix cement and Schlumberger's Squeeze-CRETE. These new slurries have a better ability to penetrate through tiny spaces.

Many studies have been carried out to investigate the rheology and variables that influence gel resistance to water flow. Grattoni et al. (2001) carried out a series of experiments to investigate the relationship between polymer gel parameters (such as gel strength and polymer concentrations) and flow behaviour. Permeability is a function of both water flow velocity and polymer concentration, they discovered. Yang et al. (2002) established a mathematical model for water flow via polymer gel-impregnated tubes. Zhang and Bai (2010) demonstrated that millimetres-sized particles produced a permeable gel pack rather than full blocking in opening fractures.

Gel rheology and extrusion properties of gels in fractures are linked, according to Seright (2003). Ganguly et al. (2012) conducted a series of studies to see how fluid leakage affected gel strength when placed in fractures. The transfer of partially produced gels in fractures was studied by Sydansk et al. (2005). Wang and Seright (2006) investigated whether employing rheology measurements to determine gel characteristics in fractures is a cost-effective alternative to extrusion experiments. Wilton and Asghari (2007) investigated how fractures could improve bulk gel placement and performance. The effect of shear on flow characteristics during the installation of sealants in fractures was examined by McCool et al. (2009).

However, none of these previous works investigate the potential application of nano-silica gel to seal cracked and fractured cements. This study will use extensive experimentation to better understand and evaluate the mechanics of nano-silica gel growth and placement in cement fractures and channels. The study will begin by assessing the impact of various parameters on sealant characteristics and rheology. The gel injection will next be evaluated using a variety of fracture and channel sizes in a series of studies. After that, several experiments will be carried out to analyse water and oil leakages using nano-silica gel.

## Experimental Investigation

The capacity of Nano-silica gel to inhibit fluid leakage through channels in the cement was investigated using an experimental setup. The use of crosslinked micro-gel in cement zonal isolation has received very little experimental attention. This research will also give petroleum engineers a basic grasp of how to use Nano-silica sealer for wellbore integrity applications. The impact of Nano-silica concentration on the gelation properties of the sealant, the sealant's capacity to block routes, and the sealant's ability to halt leakages will be investigated in this study. Rheological testing, gelation time estimations, blocking efficiency, and mechanical qualities are also part of the laboratory experiments.

## Materials

**Nano-Silica Gel** - For all studies, a commercial super-absorbent silica gel (supplied by Nouryon Pulp and Performance Chemicals Inc.) was chosen as the sealant material. The material was used exactly as it was supplied from the company, with no chemical changes. The gel was a dry white granular powder before it swelled. Chloride of sodium (NaCl). A brine solution was made using commercially available NaCl with a purity of 99.99 percent.

**Class H Cement** - All of the cement specimens used in this investigation were made with Class H cement and distilled water. Using a gas pycnometer, the specific gravity of the cement was determined to be 3.18. X-ray fluorescence spectroscopy was used to determine the chemical composition of Class-H cement (XRF)

## Material Preparation

**Preparation of Cement Paste** - A two-speed bottom-drive blender was used to mix the cement slurry at room temperature. After pouring a precise amount of water into the blender, dry cement was added at a uniform rate and mixed on low speed for around 15 seconds. The blender was then covered and the mixing continued at high speed for an additional 35 seconds (API RP 10B-2). The water/cement ratio (WCR) of the cement slurry was 0.38, as per API specification 10A for API cement Class-H. (API 2010).

**Preparation of Fractured Cement Cores -** To create the cracked cement cores, a two-inch diameter hose was filled with the cement slurry and left to harden for over 72 hours, closed on one end and open on the other. Following the hardening of the cement core, it was removed from the hose. The dried core was then gently pushed against the blade of the running tile-saw until it was chopped into two half pieces. The two parts of the core are then separated to the required fracture width using proppant or coins, as shown in Fig. 1.

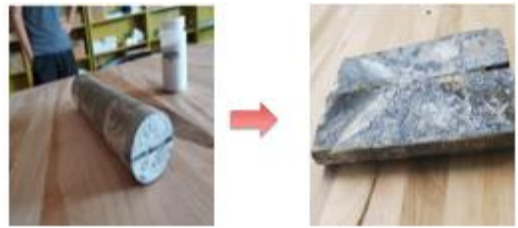

Fig 1. The Fractured Cement Core

**Gel Preparation** - A 50 percent original concentration of Nano silica solution was utilized. Brine solution was added as a gelation accelerator to aid in the development of gel in the combination. Using Nano silica concentrations ranging from 15% to 40%, salt concentration, and water, a straightforward experimental approach was employed to make silica gels at ambient temperature, as shown in Fig. 2. The Nano-silica particles were first produced at the necessary concentration. The brines with a concentration of 3% were then introduced to the Nano-silica particle and allowed to gel. Then, using an Anton Par Rheometer, the rheological and elastic properties of the resultant gel particles were measured. The Injectivity test employed NS Gel samples of 15% concentration for the gel sealing effectiveness evaluations.

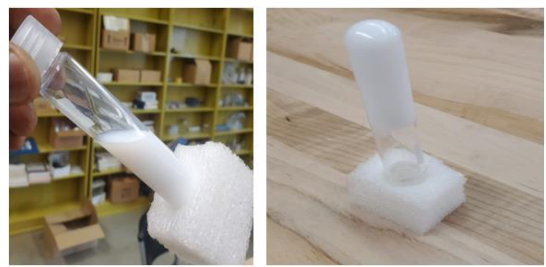

Fig 2. 15% Nano-silica gel (Before and After Gelation)

## Methodology

### Rheological Measurements

Anton Paar Modular Compact Rheometer MCR 302 Instruments with a parallel plate system (using a 100 microns gap) was used to measure the gel strength, the yield point and elastic properties. The measurements were conducted at 25 ºC to measure the gel properties of these Nano-silica gel samples.

## Experimental Procedures

Fig.3 shows the gel propagation through the cement setting. two syringe pumps, two accumulators with pistons, and a 2-inch Core holder make up the device. The fracture width was measured by inserting a 0.05-inch proppant between the two parts of the fragmented core. After that, the two pieces of the core are taped together to produce a whole core. The proppant-filled core is then inserted into the core holder's rubber sleeve and confining pressure is applied to prevent fluid leakage around the fractured core and to mimic the confining pressure conditions. Nano-silica Gel and brine were fed into the core-holder via the accumulator. To monitor the gel performance along the fracture, pressure transducers were installed at the cell's intake, middle, and outflow.

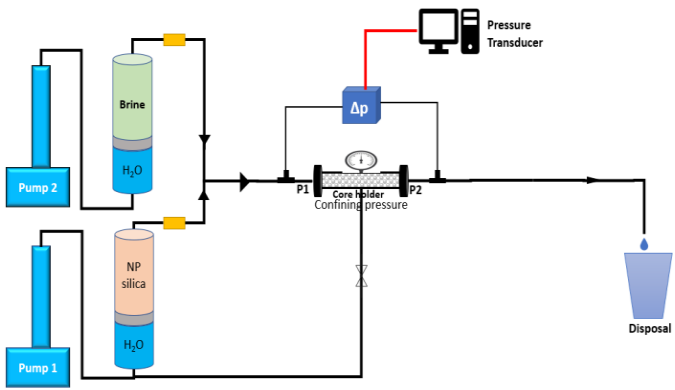

Fig 3. Experimental set up for the Injectivity Test

To simulate the circumstance when the cement was already filled with water before the sealant injection, brine was pumped into the core-holder using an accumulator. After the brine was in place, a steady flow rate of 1 cc/min of nano-silica particle gel was injected into the core-holder. The gel injection was continued until the gel was formed at the cell's exit and the injection pressure across the fracture was steady. The injection pressure was measured with pressure sensors, and the angle of gel propagation across the fracture was monitored with a high-resolution camera. To test the gel's water sealing efficiency, brine was introduced into the gel particle packed crack after it was in place. The brine was injected at a continuous flow rate of 1 cc/min through an accumulator. The brine injection was continued until the pressure of the brine injection was stabilized. Transducers were used to record pressure data for all of the trials.

## Results and Analysis

### Nano-silica Concentration effect on Gel properties

After sample preparation and testing, the observed gel generated at various nano-silica concentrations is shown in Fig. 4. The appearance of the gels demonstrates that the gel structure becomes stronger as the concentration of nano-silica particles increases. Furthermore, the gelation time reduces as the concentration increases. Gelation time is the length of time it takes for a silica solution in NP liquid form to turn into a gel. The shear stress is measured at a low shear rate after the NP Solution has sat quiescently. An NP Gel's yield point is defined as the stress at which it deforms significantly for a small increase in stretching force.

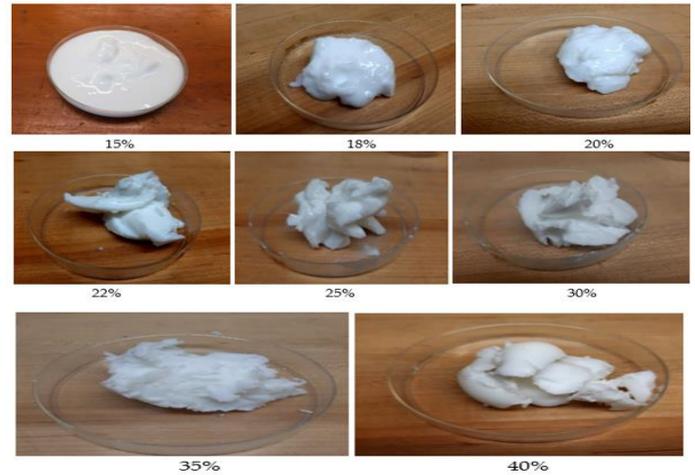

Fig 4. Nano-Silica gel at different concentration

Tab 1. Nano-Silica gel at different concentration

| NP Conc. | Gel. Time. | Gel. Str. | Yield Point |
|---|---|---|---|

| (%) | (mins) | (lb/100ft²) | (lb/100ft²) |
|---|---|---|---|
| 15 | 185 | 504 | 27.9 |
| 18 | 90 | 679 | 189 |
| 20 | 76 | 920 | 257 |
| 22 | 60 | 1546 | 452 |
| 25 | 20 | 3006 | 536.9 |
| 30 | 6 | 6160 | 1372 |
| 35 | 0 | 8484 | 8398.9 |
| 40 | 0 | 8492.6 | 8358.8 |

According to the experimental results in Tab. 1, the gel strength and yield point values increase as the nano-silica concentration increases. This is owing to the Nano-silica particle's aggressive filler network, whose rheological property increases as its concentration in the solution increases (Havet G et al, 2003). This increase in rheological feature raises the corresponding gel strength and yield point values by raising the ionic strength of the mixture solution (Gallagher & Mitchell, 2002). Furthermore, the greater the crosslinking density with the crosslinking agent, and therefore the stronger the gel strength, the higher the nano-silica concentration.

In addition, looking at the texture of the generated gel in Fig. 4 demonstrates a link between gel texture and nano-silica concentration. Increasing the nano-silica concentration accelerates the rheological property, which has a major impact on the solution's thickening and elastic properties. At a nano-silica concentration of 15%, the observed gel has a light texture in liquid form, as shown in Fig. 3, which also correlates to gel strength and yield point values of 504 100lb/ft² and 27.9 100lb/ft², respectively as seen in Fig. 5, with a prolonged gelation time of 185 minutes. The texture of the related gel thickens as the concentration of nano-silica particles in the solution increases due to the higher molecular weight caused by the inclusion of the nano-silica particle concentration in the solution. The presence of hydroxyl groups in nano-silica particles structure leads to an easy filler network in the polymer chains and caused the gel strength and yield point to increase (Zareie, C. et al. 2019).

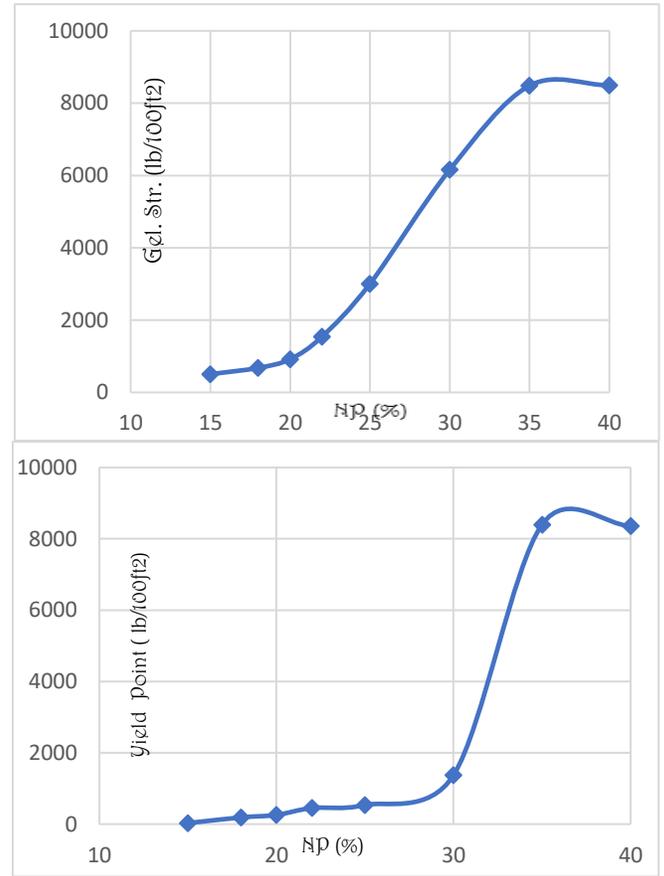

Fig 5. Gel. Strength & Yield Point at different Nano-silica concentrations

### Gel - Cement Sealing Efficiency

The Gel-Cement Plugging Efficiency setup was used to examine the gel's ability to plug cement fractures and limit water generation in this region (Fig. 3).

### Establishing a Stable Pressure Prior to Gelation

Table 2 depicts the pressure of gel injection at various flowrates. In this study, a cement fracture of 0.05 inch was used. As the flowrate increases, the pressure of the gel injection increases. The highest-pressure value of 265 psi was achieved with the highest flowrate of 4 cc/min. The injection stable pressure declined to 173, 113, and 75 psi, respectively, when the flowrate was reduced to 3, 2, and 1 cc/min. These findings show that as the flowrate increases, the pressure drop increases, following the Darcy law of flow.

To estimate the fracture conductivity during the injection test, Eqn.1 is used.

$$C_f = \frac{8.036\, q\mu\, dL}{D\, dP} \tag{1}$$

Where q (flowrate) in cc/min, μ (viscosity) in cp, $dL(core\ length)$ in inch, D (core diameter) in inch and $C_f$ (fracture conductivity) in mD-ft.

The hydraulic aperture of the fracture, B is calculated using Eqn. 2

$$B = 0.3937\left(\frac{0.00242\ Q\mu L}{W\ dP}\right)^{1/3} \quad (2)$$

Where B is measured in inch, Q in cc/min, μ is in cp, L in inch, w in inch and dP in Psi.

Tab 2. Experimental result at different flowrates

| Flowrates | Proppant | | | Dime | | |
|---|---|---|---|---|---|---|
| (cc/min) | P (psi) | $C_F$ (mD-ft) | B (inch) | P (psi) | $C_F$ (mD-ft) | B (inch) |
| 1 | 75 | 0.54 | 0.073 | 64 | 0.628 | 0.077 |
| 2 | 113 | 0.71 | 0.081 | 94 | 0.855 | 0.086 |
| 3 | 173 | 0.69 | 0.080 | 140 | 0.861 | 0.086 |
| 4 | 265 | 0.61 | 0.076 | 195 | 0.824 | 0.085 |
| **Average** | | **0.64** | **0.08** | | **0.79** | **0.08** |

From Tab. 2, the average values for fracture conductivity and hydraulic aperture are calculated as 0.64 mD-ft and 0.08inches respectively using proppant and dime of equal thickness of 0.05 inches. The purpose of using dime and proppant is to double-check the effect of proppant embedment which reduces the fracture width and conductivity.

**Establishing a Stable Pressure After Gelation (Permeability Test)**

This process involves measuring the resistance to waterflow after Gel placement. To evaluate the Nano-Silica gel blocking behavior against water, brine was introduced at a steady flowrate of 1cc/min after the gel was placed into the cement fracture. The flow resistance at various silica concentrations is shown in Figure 6. It was discovered that the gel with the greatest concentration (25%) had the highest pressure to reach a water breakthrough at 1400 psi, which dropped to around 520 psi when the Nano-Silica concentration was reduced to 13 percent. The pressure dropped once the water broke through until it reached a stable level. In comparison to gels with lower concentrations, gels with higher concentrations showed higher steady pressure. The gel's ability to plug cement fractures and manage water generation may also be controlled, with the nano-silica concentration boosting this ability.

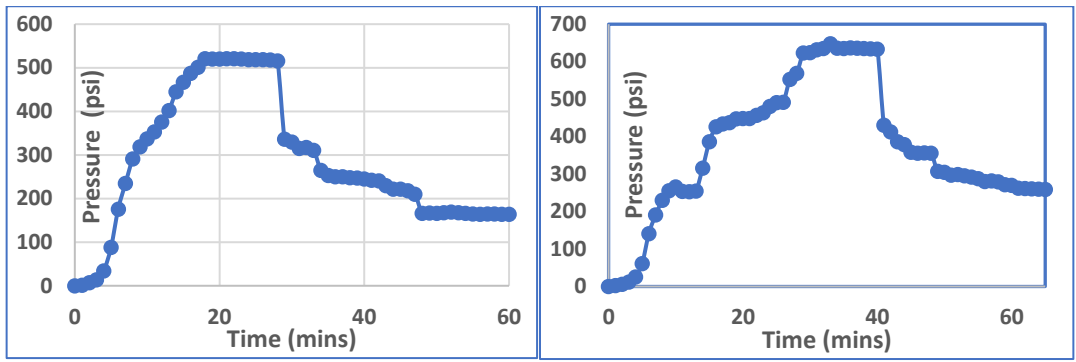

          13% Nano-silica                             15% Nano-silica

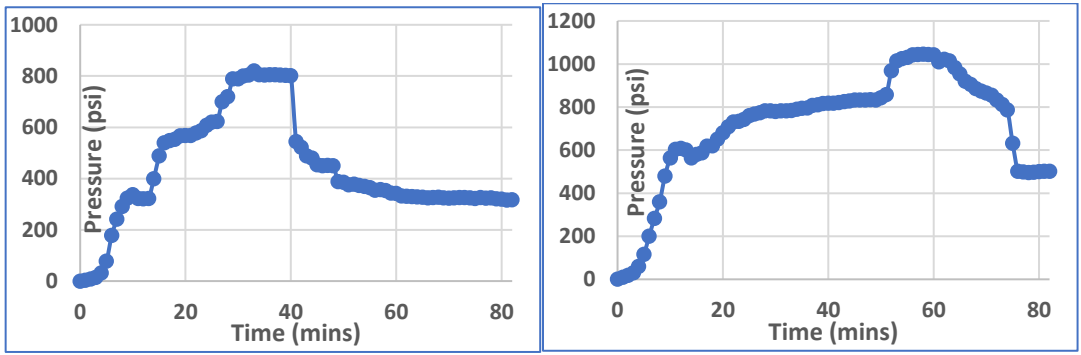

          18% Nano-silica                             21% Nano-silica

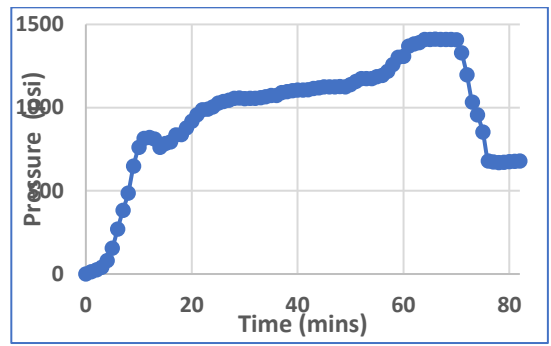

25% Nano-silica

Fig 6. Flow resistances at different Nano-silica concentrations

### 4.2.3. Water Residual Resistance Factor (FRRW)

The ratio of water phase permeability before and after particle gel treatment is known as the Water Residual Resistance Factor. The gel's plugging efficiency improves as the FRRW value rises, hence a high FRRW value is advantageous. In this study article, FRRW was determined using Eqn. 3.

$$FRRW = \frac{BRINE\ STABLE\ PRESSURE_{AFTER\ GEL}}{BRINE\ STABLE\ PRESSURE_{BEFORE\ GEL}} \quad (3)$$

The percentage of permeability reduction can be computed from Equation 4 to reflect the Nano-silica gel sealing efficiency (E) (Imqam et al. 2015)

$$Efficiency = [1-(\frac{1}{FRRW})]*100\% \quad (4)$$

Tab 3. Sealing Efficiency at different flowrates

| NS Conc. % | Pressure before Gel (psi) | Pressure After Gel (psi) | FRRW | Efficiency (%) |
|---|---|---|---|---|
| 0 | 75 | 75 | 1 | 0% |
| 13 | 75 | 520 | 6.933333 | 86% |
| 15 | 75 | 635 | 8.466667 | 88% |
| 18 | 75 | 805 | 10.73333 | 91% |
| 21 | 75 | 1044 | 13.92 | 93% |
| 25 | 75 | 1400 | 18.66667 | 95% |

Tab. 3 demonstrates that as nano-silica concentration rises, FRRW rises as well. This suggests that raising the nano-silica concentration makes water shut-off applications more successful. FRRW rose with increasing nano-silica concentration; for example, FRRW of 13 percent nano-silica gel was 6.93, while FRRW of 25 percent nano-silica gel was 18.667.

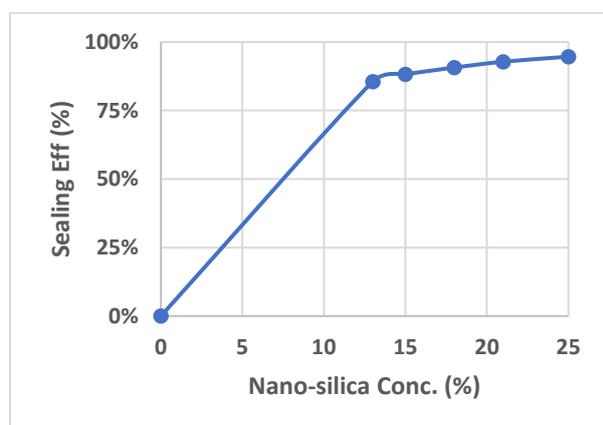

Fig 7. Sealing Efficiency at different Nano-silica concentrations

When a nano-silica gel was used as a sealant agent, the results showed that the sealing efficiency rose as the concentration of the gel increased. For all nano-silica concentrations, the sealing efficiency varies from 86 percent to 95 percent. A summary of nano-silica concentrations, FRRW, and sealing efficiency may be found in Table 3.

### Conclusions

Several findings were produced in this study by investigating the impact of infusing nano-silica gel into cement fractures. Gel rheological measurements, gel strength, gel transportation testing, and gel sealing efficiency trials were used to arrive at these conclusions. The most important conclusions are listed below.

1. Smaller fracture widths and crack characteristics suggest that nano-silica gel particles have appropriate injectivity. Their water leakage plugging performance, on the other hand, is up to 1400 psi. The gel strength and concentration might be changed to control the sealing pressure.

2. Gel with a 25 percent nano-silica concentration had the highest sealing efficiency of 95 percent, while gel with a

13 percent nano-silica concentration had the lowest sealing efficiency of 86 percent.

3. Gel sealant material selection is influenced not only by plugging efficiency but also by gel injectivity. When compared to a low-strength gel, the high-strength gel had superior blocking efficiency but poorer injectivity, and vice versa.

4. The sealing capacity of the selected nano-silica gel increases as the concentration increases. This is attributed to the fact that the concentration of nano-silica is directly proportional to its gel strength values, which also govern the sealing capacity.

## Acknowledgement

The author would like to express his gratitude to the University of Louisiana at Lafayette (ULL) for providing him with a scholarship. Thank you also to the Department of Energy in the United States for supporting the research.

## Nomenclature

% - Percent

API – America Petroleum Institute

B – Hydraulic Aperture

cc/min – Cubic centimeter per minute

$C_F$ – Fracture Conductivity

D – Darcy

FRRW – Water residual resistance factor

ft – Feet

in – Inch

Inc. – International

MCR – Modular Compact Rheometer

mD – Millidarcy

mD-ft – Millidarcy feet

NaCl – Sodium Chloride

NP – Nano-particle

NS – Nano-silica

PSI – Pounds per square inch

ºC – Degree Centigrade

WCR – Water – Cement ratio

XRF – X-ray Fluorescence